\documentclass{IATEDarticle}

\usepackage{graphicx}
\usepackage{hyperref} 
\usepackage{multirow}
\usepackage{lscape}
\usepackage{wrapfig}
\usepackage{float}
\newcommand\T{\rule{0pt}{2.6ex}}       
\newcommand\B{\rule[-1.2ex]{0pt}{0pt}} 

\begin{document}

\title{Addressing the ``Leaky Pipeline'': A Review and Categorisation of Actions to Recruit and Retain Women in Computing Education}

\author{Alina Berry\textsuperscript{1}, Susan McKeever\textsuperscript{2}, Brenda Murphy\textsuperscript{3}, Sarah Jane Delany\textsuperscript{4}}

\IATEDaffiliation{
\textsuperscript{1}Technological University Dublin (IRELAND) \\
\textsuperscript{2}Technological University Dublin (IRELAND)\\
\textsuperscript{3}University of Malta (MALTA)\\
\textsuperscript{4}Technological University Dublin (IRELAND)\\
}

\maketitle

\begin{abstract}

Gender imbalance in computing education is a well-known issue around the world. For example, in the UK and Ireland, less than 20\% of the student population in computer science, ICT and related disciplines are women. Similar figures are seen in the labour force in the field across the EU. The term ``leaky pipeline'' is often used to describe the lack of retention of women before they progress to senior roles. Numerous initiatives have targeted the problem of the leaky pipeline in recent decades.

This paper provides a comprehensive review of initiatives related to techniques used to boost recruitment and improve retention among women in undergraduate computer science and computing courses in higher educational institutions. The review covers 350 publications from both academic sources and grey literature sources including governmental guidance, white papers and non-academic reports. It also includes sources in languages other than English. 

The primary aim was to identify interventions or initiatives (which we have called ``actions'') that have shown some effectiveness. A secondary objective was to structure our findings as a categorisation, in order to enable future action discussion, comparison and planning. 

A particular challenge faced in a significant portion of the work reviewed was the lack of evaluation: i.e. the assessment of the direct relationship between the initiatives undertaken and the outcomes on retention or recruitment. There are only a limited number of studies that include a control group and these tend to focus on one particular intervention or action. In addition often the work presents a number of actions that were implemented and it is difficult to determine which action produced most impact. Considering these challenges, actions were identified that had some level of evaluation with positive impact, including where the evaluation was by measuring feedback.

The actions were categorised into four groups: Policy, Pedagogy, Influence \& Support and Promotion \& Engagement. \textit{Policy} actions require support and potentially structural change and resources at organisation level. This can be at a department or school level within a higher level institution, and not necessarily just at the higher institution level. \textit{Pedagogy} related actions are initiatives that are related to the teaching of computer science and technology in terms of curriculum, module delivery and assessment practice. The \textit{Influence and Support} category includes actions associated with ways to influence women to choose computing at third level and once enrolled to support and encourage them to stay in the field. Finally, \textit{Promotion and Engagement} actions are initiatives to promote computer science and technology based courses and involve engagement and outreach with external stakeholders such as industry, communities and schools.

We present our categorisation, identifying the literature related to actions under each category and sub-category. We discuss the challenges with evaluating the direct impact of actions and outline how this work leads towards the next phase of our work – a toolkit of actions to promote retention and recruitment of women in computing based undergraduate courses.

This work will be of interest to third level institutions with STEM faculties, gender-balance policy makers, technical industry players, or any stakeholder in the field of STEM who wishes to understand and implement solutions to the imbalance of women in computing education and beyond.

\end{abstract}

\section{INTRODUCTION}

Despite a significant amount of research and practical initiatives that have taken place to stimulate gender balance in computing fields\footnote{In this paper all disciplines including Informatics, IT, ICT, hybrid courses and related fields that have a significant programming component in their curriculum are referred to as Computing, Computer Science or ICT.}, the problem still persists. In 2021 only 19.1\% of  ICT specialists working in the EU were women \cite{Eurostat2020a}. 
With regard to tertiary education,there are less than men in ICT disciplines. For Ireland, the latest available report for new entrants in 2017/18 \cite{HEA} shows that on average nationally just 14.8\% of ICT related course enrolments were women. A recent UK report \cite{STEMWomen} indicates that just 19\% of computer science students were women.
EU statistics show a similar pattern. In 2018 girls and women accounted for only 17\% in upper secondary and tertiary ICT education on average across all countries \cite{Eurostat2020b}.

Without enough women in computing education there will be limited chances to increase women's participation in relevant careers further down the pipeline. The metaphor ``leaky pipeline'' is commonly used to describe the lack of retention of women in STEM fields before they continue to more senior roles \cite{goulden2011}.

Some of the frequently mentioned reasons for the leaky pipeline include stereotypes about computing being a male domain \cite{cheryan2013stereotypical}, cultural influence \cite{frieze2019computer}, a lack of a sense of belonging \cite{aelenei2020academic} and family and caring matters \cite{ayre2014family}.

Anderson et al. \cite{anderson2017can} discuss that these reasons are not universal and do not provide a solid foundation for extensive replicable initiatives for recruitment and retention of women in computer science. Nonetheless, there has been a large number of success stories reported in different parts of the world \cite{campLiebeslattery, frieze2019computer, lagesen2007strength, mckeever2021addressing, onyema2021effect, siegeris9mehr}. Hence, we find it important to collect and discuss effective initiatives that worked in a range of settings.
This paper aims to produce a comprehensive review and categorisation of initiatives used around the world. It focuses on those that showed impact on recruitment and retention of female students in technology related courses. A significant challenge faced in tackling the problem of gender imbalance lies in the assessment of initiatives and the interpretation of results. We discuss the common techniques and measures used in evaluation. We also note any observed limitations of positive impact on certain groups of students.

The rest of this paper is structured as follows: Section \ref{Approach} outlines our sources of literature, and our methodology.  
Section \ref{Actions} outlines our proposed categorisation of actions, which is presented visually, and discusses the actions in detail. Section \ref{discussion} discusses the challenges associated with the measurement of impact and discusses the methods that are currently used. Section \ref{conclusions} concludes the paper.

\section{APPROACH}\label{Approach}

We focused our review on the technical higher education stage and successful recruitment and retention techniques relevant to gender minorities, primarily women, were targeted.
We identified the relevant sources by using online search engines and digital databases, by conducting manual searches and by inspecting grey literature. Additionally, a search in languages other than English was performed, including German, Spanish, Ukrainian and Russian.

We used online search engines such as Google Search, Google Scholar, Microsoft Bing, Twitter Search, multi-databases platforms such as EBSCOhost, ProQuest and TU Dublin library catalogue. Databases such as ACM Digital Library and Emerald Insight were searched as well. Keyword combinations in relevant languages such as, “recruitment of women in computing”, “effective techniques to recruit and retain women in STEM higher education”, “women in technology initiatives”, “recruitment and retention in computer science”, “female gender recruitment in technology”, “interventions to attract women into computing”, “female student recruitment in computing”, “initiatives for women in computing” were dealt with. 

Manual searches were then done to look for additional relevant materials such as browsing through earlier or later volumes of relevant journals, and searching for additional publications of researchers specialising in the field. 

We identified three different sources of literature among over 350 relevant publications. First are research papers from journals and academic conferences, as well as academic reports.  We included academically published studies and interventions, which focused on addressing the problem of gender imbalance in computer science education. This group offered the largest pool of literature as compared to the other two. Nonetheless, many sources focused on theoretical discussions or research based recommendations, as opposed to evidence based studies.

Second, given that many gender balance in STEM actions are not formally reported in peer-reviewed publications, we considered material outside of the academic collections to identify practical advice and initiatives. These included submission papers for international awards \cite{minerva}; web portals of initiatives aiming to increase gender diversity in computing, such as Ada in Norway \cite{ADAproject}, Czechitas in Czechia \cite{Czechitas}, INGENIC in Ireland \cite{Ingenic}, Ladies in Informatics in Slovenia \cite{LadiesInInformatics} or MIT Women's Technology Program in the United States. 


The third source covers governmental/institutional guidance - e.g., reports by international organisations and recommended strategies. Depending on their focus, authors dealt with the improvement of existing practices at a policy level \cite{lagesen2007strength, siegeris9mehr}; teaching level \cite{albarakati2021rethinking, gutica2021fostering, buhnovahappe2020girl}; support level by role models/mentors \cite{frieze2019computer, tupper2010strategies}; or used outreach strategies to attract students into the field \cite{modekurty2014c, roden2013growing}.



Two categories of approaches on how to improve gender balance in computer science education have emerged from the literature: \textit{Frameworks} and \textit{Actions}.
Frameworks offer a structure or strategy at a high level to promote and improve gender balance. Often they are presented as a process or step-by-step guide which typically involves key tasks such as self-assessment, planning, roles and responsibilities, recruitment and monitoring. They can offer practical advice on strategies and can include case studies. They can also be called workbooks, models, toolkits, toolboxes, or tools by their authors \cite{achenbach2018systemic, EIGE, moredapozo2020csframeworks, NCWIT2009, NCWIT2015, toolboxNtnu}.
Actions are practical initiatives that can be implemented and evaluated individually. We focus on actions with some support in the literature for having had successful impact. 

The following section introduces our categorisation of actions and discusses initiatives that have demonstrated effectiveness in either encouraging women into technology-focused courses or assisting in the retention of women on these courses. 
In limited cases, relevant initiatives have been considered from disciplines such as general STEM \cite{benavent2020girls4stem, breda2020gender, nash2017understanding}; engineering \cite{botella2019gender, dennehy2017female, marley2019promoting, xu2017getting}; or where women were not the focus but impact was significant \cite{archer2017impact, ott2014explorations, ott2018impact, sobral2021flipped, roden2013growing, gutica2018improving, galloway2014use}.

\section{CATEGORISATION OF ACTIONS} \label{Actions}

\begin{figure}[H]
  \centering
  \includegraphics[scale=0.9]{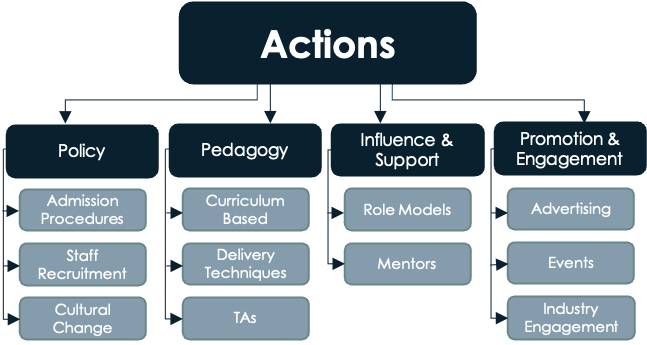}
 \caption{\textit{Categorisation of actions.}}
  \label{fig:Actions}
\end{figure}

Our categorisation of actions, displayed in Fig.~\ref{fig:Actions}, consists of four groups: Policy, Pedagogy, Influence \& Support, and Promotion \& Engagement. Within each category, grouping or subcategories of actions were identified. Note that some actions could belong to multiple categories or subcategories or require support by stakeholders from other categories.

\textbf{Policy} actions require support and potentially structural change and resources at organisation level. This can be at a department or school level within a higher level institution, and not necessarily just at the higher institution level. For example, to set a enrolment quota for women, or to recruit female faculty staff.
\textbf{Pedagogy} related actions are initiatives that are related to the teaching of computer science and technology in terms of curriculum or module delivery. These types of actions could potentially be undertaken at a module level by an academic within a school or department and do not necessarily require policy changes.
\textbf{Influence \& Support} actions are associated with ways to influence female students to choose computing at third level and once enrolled to support and encourage them to stay in the field. 
\textbf{Promotion \& Engagement} actions are initiatives to promote computer science and technology based courses and can involve engagement and outreach with external stakeholders such as industry, communities, and schools.
Table \ref{tab:table1} shows the papers with actions identified as having some level of impact in the recruitment and retention of women in computing course grouped by category and sub-category.

\begin{table}[h!]
    \centering
     \caption{Papers which has shown some positive impact from actions to increase recruitment and retention of women in computing based courses grouped under category and sub-category and identifying the kind of evaluation used to measure the level of impact.}
    \begin{tabular}{lp{1.9cm}p{3.6cm}| p{3.5cm}| p{2.5cm}l}

\hline \hline   
   & &  \multicolumn{3}{c}{\textbf{Evaluation}} \T \B\\ 
\cline{3-6}
   {\textbf{Category}} & {\textbf{Sub-Category}} & & &  \\
  &  & Enrolment \& &  Feedback &  Other \\ 
  &  &  Retention Rates &   &  \B \\ 

\hline 

 & Admission Procedures & \cite{frieze2019computer}, \cite{wu2018expanding}, \cite{lagesen2007strength}, \cite{siegeris9mehr}, \cite{newsminerva2021Bremen}   & \cite{newsminerva2021Bremen}   & ... \T \B \\

 \cline{2-6}
 \textbf{Policy} & Staff  & \cite{xu2017getting}, \cite{lagesen2021inclusion}, \cite{mckeever2021addressing}  & ...  & ... \T \\
 & Recruitment &  & &  \B \\

 \cline{2-6}
 & Cultural Change & \cite{frieze2019computer}, \cite{siegeris9mehr}, 
  \cite{botella2019gender} &
 \cite{siegeris9mehr}  & ...\T \B \\

 \hline

& Curriculum  Based & \cite{mckeever2021addressing}, \cite{siegeris9mehr}, \cite{alvarado2012increasing}, \cite{brodley2022},  \cite{frieze2019computer}, \cite{roden2013growing}, 
\cite{newsminerva2021Bremen} &
 \cite{newsminerva2021Bremen}, \cite{siegeris9mehr},  \cite{carbonaro2010computer}, \cite{ryooMargolis2013}  & \cite{sigurdhardottir2019empowering} \T \B \\
 \cline{2-6}
 
 \textbf{Pedagogy} & Delivery Techniques & \cite{latulipe2018longitudinal},  \cite{albarakati2021rethinking}, \cite{minervaCopenhagen},  \cite{buhnova2019assisting}, \cite{moskal2004evaluating}, \cite{alvarado2012increasing},  \cite{mckeever2021addressing},  \cite{minerva2018bamberg}, \cite{mcdowell2006pair}, \cite{newsminerva2021Bremen},  \cite{minerva2018bamberg}  
 
 & \cite{buhnova2019assisting},  \cite{buhnovahappe2020girl},  \cite{marley2019promoting}, \cite{kamberi2015}
 \cite{nash2017understanding}, \cite{gutica2018improving}, \cite{carbonaro2010computer}, \cite{newsminerva2021Bremen}, \cite{minerva2018bamberg}, \cite{mcdowell2006pair},  \cite{miller2020examination}
 
 & \cite{albarakati2021rethinking}, 
 \cite{rahman2018leveraging},
 \cite{moskal2004evaluating},
 \cite{gutica2021fostering},
 \cite{modekurty2014c},
  \cite{fisk2021increasing},
 \cite{archer2017impact},
 \cite{modekurty2014c},
 \cite{mcdowell2006pair} 
 & 
 \B \T \T \\
 
 \cline{2-6}
 
 & TAs & 
 \cite{mckeever2021addressing},  \cite{minerva2018bamberg}, \cite{albarakati2021rethinking}
  & \cite{minerva2018bamberg}  & \cite{albarakati2021rethinking}  
 \T \B \\ 
 \hline
 
 &  Role &  \cite{minervaZagreb}, \cite{wu2018expanding},  \cite{minervaCopenhagen}, \cite{minerva2018bamberg}, 
 & \cite{yates2021female}, \cite{gutica2021fostering},  \cite{minerva2018bamberg}  
 & \cite{gutica2021fostering}, \cite{breda2020gender} \T \\
  
\textbf{Influence \&} & Models &  \cite{frieze2019computer}, \cite{xu2017getting}  &    &  \B  \\ 
 \cline{2-6}
 \textbf{Support} & Mentors & \cite{minervaCopenhagen},  \cite{minervaZagreb}, \cite{alvarado2012increasing}, \cite{tupper2010strategies}, \cite{dennehy2017female}, 
 & \cite{modekurty2014c}, \cite{doerschuk2004research}
 & \cite{dennehy2017female}, \cite{modekurty2014c} \T \B\\
 \hline

 & Advertising & \cite{lagesen2007strength}, \cite{minervaZagreb},  \cite{xu2017getting}, \cite{minervaCopenhagen},  
& \cite{buhnova2019assisting}  &  \T \\
\textbf{Promotion} & & \cite{roden2013growing}  &  &   \\
 \cline{2-6}
 
  \textbf{\& Engagement } & Events  &   \cite{minerva2018bamberg}, \cite{minervaCopenhagen}, \cite{minervaRadboud}, \cite{lagesen2007strength}, \cite{siegeris9mehr}, \cite{alvarado2012increasing}, \cite{guzdial2014georgia}, \cite{mckeever2021addressing}, \cite{lagesen2021inclusion},  \cite{buhnova2019assisting}, 
  \cite{xu2017getting}, \cite{rizvi2012scratch},  \cite{ott2018impact}, \cite{tupper2010strategies}
 & \cite{minerva2018bamberg}, \cite{siegeris9mehr},  \cite{minervaWestBohemia}, \cite{gutica2021fostering}, 
 \cite{drobnis2010}, \cite{mckeever2021addressing}, \cite{minervaRadboud}
 & \cite{egan2010recruitment}, \cite{drobnis2010}, \cite{gutica2021fostering}, \cite{tupper2010strategies}, \cite{rursch2009adventures}
 & \T \B\\
 
 \cline{2-6}
 & Industry  & \cite{ott2014explorations}, 
 \cite{fryling2018catch}, \cite{newsminerva2021Bremen},  
 & \cite{newsminerva2021Bremen}, \cite{ott2014explorations},  \cite{pantic2020retention}, \cite{fryling2018catch},  &  \cite{ott2014explorations}  \T \\
 & Engagement & \cite{mckeever2021addressing} & \cite{buhnovahappe2020girl} & \B \\ 
 
 \hline
\hline
    \end{tabular}
    \label{tab:table1}
\end{table}

\subsection{Policy}

Policy actions can involve change to processes or policies at institutional or departmental level. 
These will generally require high level of support from leadership posts and may require significant resources to be assigned in order to implement. 

A significant number of actions in this category are associated with \textbf{Admission Procedures}. Some examples include changing the entry requirements to remove the requirement for computing or programming background \cite{frieze2019computer} and facilitating students moving courses more easily \cite{schmid2015neigung, wu2018expanding}.  
More radical actions have been successful in Norway where Norway University of Science and Technology (NTNU) instigated a quota for students and permitted reduced points for female candidates in 2007 \cite{lagesen2007strength}. 
There are women-only computer science courses emerging now \cite{newsminerva2021Bremen, siegeris9mehr}.

There are also examples of actions that can be beneficial for female \textbf{Staff Recruitment} including adjusting job descriptions to make them more appealing for women to apply, or targeting specific individuals though personal contact and setting targets for hiring \cite{xu2017getting}. 
Some positive experience has been reported after appointing a dedicated champion/lead of the project \cite{mckeever2021addressing, lagesen2021inclusion}.

The need for \textbf{{Cultural Change}}
and the appropriate management of this change at institutional level are identified as important steps towards gender equality in technical disciplines \cite{frieze2019computer, mckeever2021addressing, xu2017getting}. Practical examples are the recognition of outstanding female students with awards \cite{botella2019gender}; organisations for faculty members and students together that provide institutional support and funding for initiatives \cite{frieze2019computer}; and providing a guaranteed family friendly study schedule for students with children or carer's responsibilities \cite{siegeris9mehr}.

\subsection{Pedagogy}

Pedagogical practices include actions that deal with the curriculum and teaching activities, and can be typically undertaken by the academic staff. In some cases support from a higher level may be needed. Some relatively simple \textbf{{Curriculum Based}}
changes have been effective such as revising module sequences to balance practical and theory modules evenly in year one \cite{mckeever2021addressing}, or a non-obvious hardware module that ends with an installation party \cite{newsminerva2021Bremen}. Hybrid courses including modules from other disciplines such as business \cite{siegeris9mehr}, biology \cite{alvarado2012increasing}, a foreign language \cite{mckeever2021addressing}, or giving options to choose from many combined majors\cite{brodley2022} have shown good success at being more attractive to female students. 
Frieze \& Quesenberry \cite{frieze2019computer} report from their experience at Carnegie Mellon University that curriculum change needs to benefit all, and not only women, when implemented.

Some actions involving game development collected positive feedback including a gender neutral interactive adventure game authoring for high school students \cite{carbonaro2010computer}, a 2 hour workshop to introduce women to creative IT including game development \cite{sigurdhardottir2019empowering}, and creating video games that are personal and meaningful to women and tell their own stories \cite{ryooMargolis2013}. There is evidence that game degrees contributed to increased enrolment for all students \cite{roden2013growing}. Previous experience seems to be particularly important for women when deciding in favour of creative technical subjects \cite{sigurdhardottir2019empowering}.

Induction is known to be an important event for new students particularly to stress the value of incorporating the key message of participation and attendance as part of the induction \cite{mckeever2021addressing}. A mandatory induction  social event is recommended for new students as a possible benefit to later retention \cite{talton2006scavenger}.
Focusing on \textbf{Delivery Techniques} also has had positive impact. The flipped classroom model is promoted as a promising approach for teaching computer science \cite{sobral2021flipped} and it has been used successfully in a lot of research into encouraging female students in computer science \cite{albarakati2021rethinking, latulipe2018longitudinal, onyema2021effect}.
It has also been shown that the flipped classroom model positively influences the academic accomplishments of female students \cite{onyema2021effect} and the retention of women in computing classes \cite{latulipe2018longitudinal}. Conversely, it does not seem to benefit men as much as women \cite{latulipe2018longitudinal, onyema2021effect} as well as less motivated students in general \cite{sobral2021flipped}. 
In some cases it was helpful to female students to be provided with e-learning courses repeating the material from the education plan \cite{buhnova2019assisting} or provide all the study material upfront and a video recording after the class \cite{buhnovahappe2020girl}.
Active learning strategies are commonly mentioned as beneficial for female students alongside the flipped classroom teaching model. Specific techniques that had impact include pair programming \cite{albarakati2021rethinking, gutica2021fostering, mcdowell2006pair, miller2020examination}, live coding \cite{albarakati2021rethinking, minervaCopenhagen}, practice-based learning \cite{marley2019promoting}, particularly in a game environment \cite{kamberi2015}, and peer instruction \cite{albarakati2021rethinking, buhnovahappe2020girl, latulipe2018longitudinal}.
Approaches such as project based work \cite{nash2017understanding, modekurty2014c} and encouraging a culture of discussion \cite{nash2017understanding}  have also shown impact. 
Gutica \cite{gutica2018improving} found that students were engaged and motivated, demonstrated improved communication and team work skills through use of a large agile software development project. Additionally, agile software development methods are, according to Frieze et al. \cite{frieze2006culture}, a better match to women's style of management.
Visual programming has shown to be beneficial for increasing female participation in computer science \cite{carbonaro2010computer, moskal2004evaluating, rahman2018leveraging}. It has also been especially effective for retention in introductory courses \cite{rizvi2012scratch}.

Keeping female students with similar programming experience together in a class and not mixing novices with more experienced programmers was found to have a strong impact on the students' perception of difficulty \cite{archer2017impact}. The level of programming experience was determined using a mandatory assessment as part of the entry procedures. Other authors support the idea to split a class \cite{alvarado2012increasing, buhnovahappe2020girl} or a practical lab \cite{albarakati2021rethinking} by experience as part of their overall efforts to improve participation and retention among women.

Personalisation is discussed by a number of scholars as effective tools to encourage the retention of female students. These include, for instance, sending personalised supportive emails with exam results which increased the retention of top performing students \cite{fisk2021increasing}; although sending the bottom-performing students emails with resources that might assist them had no effect. 
Knowing and addressing all students by name in class \cite{newsminerva2021Bremen}; keeping focus on less experienced learners \cite{buhnovahappe2020girl}; following up with individual students on any lack of attendance \cite{mckeever2021addressing}; designing bonus assignments where appropriate \cite{buhnovahappe2020girl} or modifying class content \cite{buhnovahappe2020girl} are other practical examples of personalised actions focused on female students.
To reduce the level of isolation for some students, seating can be shuffled randomly at the beginning of each class \cite{albarakati2021rethinking}, or each female student placed in groups with at least one other woman \cite{mckeever2021addressing}. 

There is evidence that in-class actions can encourage female students' participation. These include designing problem-based lessons that challenge and interest students \cite{nash2017understanding}; encouraging a culture of discussion \cite{buhnovahappe2020girl} or seminars where students of all genders can reflect on their own experiences \cite{minerva2018bamberg}.

\textbf{Teaching Assistants} (TAs) play their part in increasing students' confidence. The use of undergraduate students \cite{mckeever2021addressing, benavent2020girls4stem} and postgraduate students \cite{minerva2018bamberg} as TAs have been part of the efforts that helped female retention. 
It was shown that a diverse TA team in terms of gender and race could help in influencing the retention of women \cite{albarakati2021rethinking}. 
In addition to providing help and increasing students' confidence, female lab assistants and TAs also serve as role models to younger female students \cite{minerva2018bamberg}.

\subsection{Influence \& Support}

The gendered nature of computer science stereotypes can deter women from becoming interested in the field \cite{cheryan2013stereotypical}. There are some simple ways to counteract this such as avoiding objects that enforce the stereotypical computer science environment (Star Trek posters, video games) \cite{cheryan2009ambient}. There is a lot of evidence in the literature that role models and mentoring have a strong impact in this area. Both are discussed below.

\textbf{Role Models} have been shown to have a positive impact on encouraging and inspiring women to pursue computer science at different stages throughout the academic lifecycle. At recruitment, female students or faculty members can serve as ambassadors for annual fairs \cite{minervaZagreb} or school visits \cite{wu2018expanding}. Female computer science students from a UK university reported that their enablers were often male family members from the industry, as well as teachers in school \cite{yates2021female}.
Breda et al. \cite{breda2020gender} observed in a study with top performing female school students in France that the most effective role models were industry experts who did not emphasize gender disparities but instead delivered a positive image of their fields. Gutica \cite{gutica2021fostering} found that involving undergraduate role models of all genders for school students can improve the confidence of women in coding. University students have also been engaged to become digital ambassadors on social media \cite{minervaCopenhagen}. 
Regular interaction with role models can be supportive for existing students as well, for instance in the form of informal meetings at the faculty \cite{frieze2019computer, minerva2018bamberg}. Increasing female faculty provides more female role models for students also \cite{xu2017getting}.

To target future students of computer science, \textbf{Mentoring} programs for secondary schools that link with existing CS students can be organised \cite{minerva2018bamberg}. At college level, mentoring by more senior students helps to overcome programming difficulties \cite{minervaCopenhagen, minervaZagreb}. A focus of mentoring the more inexperienced learners in their projects is also recommended
 \cite{alvarado2012increasing}. 
Peer mentoring is discussed by a number of researchers as helpful to female students \cite{doerschuk2004research, modekurty2014c}. Matching students and mentors based on common interests outside of education has also shown success \cite{tupper2010strategies}.
While female mentors can help with confidence, motivation and increased retention \cite{dennehy2017female}, some authors suggest that engaging both female and male mentors can be effective \cite{minervaUtrecht, tupper2010strategies}. There are examples of mentoring networks with male mentors alongside the female in sub-fields, such as programming languages research \cite{Sigplan} and music information retrieval \cite{Wimir}.

\subsection{Promotion \& Engagement}

This category includes actions related to advertisement and promotion of computer science to prospective female students and engagement with external stakeholders such as schools, other universities and industry.

Various strategies have been  used for \textbf{Advertising} computer science courses with the aim of encouraging more female applicants. These include a national cinema campaign \cite{lagesen2007strength}; brochures \cite{lagesen2007strength}, in particular with a focus on gender neutral design and presentation \cite{minervaZagreb}; a dedicated career guidance portal \cite{buhnova2019assisting}; online advertisement \cite{xu2017getting} and promotion through videos, articles and examples of projects which are designed  to explain software development \cite{minervaCopenhagen}. 
Promoting state of the art equipment in the computer labs has been effective \cite{roden2013growing} as has  
promoting good gender ratios \cite{xu2017getting}. However, the chicken and egg problem with gender ratios has been acknowledged  and it is easier to attract women as female enrolment rises \cite{xu2017getting}.

A range of outreach \textbf{Events} are recommended as effective ways to encourage women applicants into computing. Face-to-face information events, such as educational fairs in secondary schools \cite{minervaCopenhagen}, open days which focus specifically on the science \cite{minerva2018bamberg}, or women's days which focus on attracting female applicants \cite{lagesen2007strength, siegeris9mehr, minervaRadboud}. 
There are many examples of outreach introductory classes to teach programming and coding to school girls \cite{ott2018impact, minervaCopenhagen, alvarado2012increasing, rizvi2012scratch, drobnis2010}. 
A mid-year female technology camp is one of the most successful outreach activities at the University of Copenhagen, with 30\% student application rate afterwards \cite{minervaCopenhagen}. Other examples include summer camps with activities such as animation, virtual reality or robotics, which increased enrolment \cite{guzdial2014georgia, minervaWestBohemia}. 
There are various way to encourage participation at the events. 
Students were asked to bring a friend they are comfortable to work with \cite{gutica2021fostering}, some have mandated 50/50 gender balance \cite{mckeever2021addressing} and others created an all-girls' environment in a summer school dedicated to computer science \cite{drobnis2010}. Others selected students to attend based on their mathematical abilities and motivation letter \cite{lagesen2021inclusion}. 
Workshops can include research activities as well \cite{minerva2018bamberg}. 
There are also examples of weekend workshops for more mature women who might be willing to change careers \cite{buhnova2019assisting}. 
The use of female faculty and students is important in outreach activities \cite{minerva2018bamberg, minervaCopenhagen}
Other types of events including external engagement which have helped in contributing to increased female student retention include Grace Hopper Celebration of Women in Computing \cite{alvarado2014}, career days with guest speakers from industry \cite{tupper2010strategies} and industry networking events to facilitate the students meeting businesses \cite{lagesen2021inclusion}.

Contests are another way of reaching out to prospective students \cite{egan2010recruitment, rursch2009adventures, xu2017getting}. 
Massachusetts Institute of Technology encourages female students to participate in the annual math competition and offers the winner the world's largest remuneration \cite{xu2017getting}.
An innovative way to attract talented math students from secondary schools is a one day competition that takes place on a school day during school hours \cite{egan2010recruitment}. While this activity was well received by students, findings show that it also improved the perception and awareness of computer science amongst school teachers.
Rewards can also be successful. 
To motivate students to enrol and to stay Tupper et al. \cite{tupper2010strategies}  discuss reimbursement of course fees from a scholarship fund upon successful completion of technical modules.

\textbf{Engagement with Industry} has also demonstrated impact. Some examples of actions are inviting industry partners to university course teaching \cite{newsminerva2021Bremen, ott2014explorations}, co-organising workshops \cite{minervaUtrecht} or coding clubs supervised by industry partners \cite{buhnova2019assisting}. 
Industry work placement has been shown to increase enterprise and employability skills and commercial awareness for IT students \cite{galloway2014use}. 
There is evidence that work experience is attractive to female students and can encourage ``sticking with the programme'' \cite{pantic2020retention}. Some different approaches to placement which have had impact include international placements \cite{mckeever2021addressing} and the dual programme of the women-only course at Hochschule Bremen \cite{newsminerva2021Bremen}, where students are taking internships throughout their course.
Fryling et al. \cite{fryling2018catch} found that a higher percentage of women undertook internships and had good retention and recruitment from non-computing majors.

\section{DISCUSSION}
\label{discussion}

One of the challenges with any initiative to encourage more female participation in computing is measuring the impact and knowing what actually works and results in good impact.  
A number of researchers recognise the challenge of assessing the direct relationship between the efforts taken and the outcomes on recruitment or retention \cite{lagesen2021inclusion, mckeever2021addressing, wu2018expanding}. It is difficult to directly relate a once-off event, even with positive feedback, to later recruitment or retention success. 
There are only a limited number of studies that include a control group and these studies tend to focus on one particular intervention or action \cite{kamberi2015, moskal2004evaluating, onyema2021effect, rizvi2012scratch}.
It is also difficult to determine which action produced most impact where multiple actions have been implemented \cite{albarakati2021rethinking, mckeever2021addressing}.
A number of researchers focus on measuring feedback, rather than impact. The level of satisfaction is a frequently mentioned form of feedback. It can be, for instance, satisfaction in the course of study and supervision \cite{minerva2018bamberg}, appreciation for the newly implemented way of teaching \cite{gutica2018improving}, or positive feedback from an event \cite{mckeever2021addressing}. The usefulness of measuring satisfaction really depends on the significance of the question being answered by measuring satisfaction levels.  

Most evaluation of impact is one of two types, empirical evaluations and assessment from experience. 
Actions with quantitative empirical evaluation, which formed the majority of our findings, include pre-planned once-off interventions or longitudinal studies which collected data and measured performance in some way. 
Those actions whose impact was based on experience were assessed more qualitatively. They could occur in teaching and learning events and be reported by observers or participants later. In some cases evidence was collected through interviews \cite{lagesen2007strength, xu2017getting}, focus groups \cite{buhnovahappe2020girl, rizvi2012scratch}, and during observation of students \cite{gutica2018improving}. 
Anecdotal evidence has also been reported \cite{alvarado2012increasing, gutica2018improving, rizvi2012scratch}. The primary way to collect the data is through surveys and questionnaires \cite{albarakati2021rethinking, mckeever2021addressing, siegeris9mehr, tupper2010strategies, albarakati2021rethinking, marley2019promoting}.

A variety of performance measures have been reported when evaluating the impact of actions. The most obvious ones are to directly measure female student enrolment rates \cite{minervaZagreb, wu2018expanding} and retention rates \cite{albarakati2021rethinking, lagesen2021inclusion, latulipe2018longitudinal, mckeever2021addressing}. 
However, significant changes to these statistics are unlikely in the short term, so a number of studies measure something more tangible in the short term such as students' perceived confidence in applying the skills learned \cite{gutica2021fostering, fryling2018catch, modekurty2014c} or their motivation \cite{rahman2018leveraging, dennehy2017female} or sense of belonging \cite{albarakati2021rethinking, rahman2018leveraging}. Table \ref{tab:table1} groups the papers based on evaluation types.

Studies have also reported that students' interest in computer science and their perception of computing changed following interventions \cite{carbonaro2010computer, egan2010recruitment, fryling2018catch}. 
Student performance before and after the intervention is also measured \cite{albarakati2021rethinking, ott2014explorations}.

Most of these initiatives were implemented and measured over a short period of time but there are some longitudinal studies and observations too \cite{lagesen2021inclusion, latulipe2018longitudinal}. The importance of longitudinal data collection in order to observe trends in computer science participation is stressed \cite{main2017underrepresentation}, however a challenge with longitudinal studies is the effort and resources required to undertake them.

While most efforts discussed in this paper reported some positive impact, some researchers report some ineffective interventions. These include, for instance, open days that did not provide a clear return on investment \cite{craig2008twenty}, women only labs that were not attractive to female students \cite{lagesen2007strength} and an introductory computing course for business students that did not create a desired outcome \cite{anderson2017can}.
There were also costly efforts to redefine the image computer science to be perceived as feminine which were not successful \cite{lagesen2007strength}. The message given in promotional materials was about the need for skills, as they claim, only women could offer to the computing industry - e.g., people skills or usability focus, as opposed to pure technical details.
However, these unsuccessful efforts do not necessarily mean unsuccessful actions as these actions could potentially be effective in different situations and settings. 
The factors that impact on female participation are still not clearly understood \cite{anderson2017can}. Culture also plays a role and to increase the chance of success it is recommended to design or modify initiatives with respect to the local context \cite{wu2018expanding}.
Additionally, initiatives are often implemented and evaluated by the same people. To avoid negative publicity, unsuccessful results might stay in the shadows without being published \cite{brenning2021mint}. 
In some rare cases an external evaluator was involved in the project \cite{guzdial2014georgia, tupper2010strategies}. 
The evaluator conducted the end of project survey but also monitored and reported on the progress of the initiative and organised yearly student focus group meetings, which facilitated changes based on students' recommendations \cite{tupper2010strategies}.

To address the challenge with evaluation, a framework for the assessment of STEM projects for secondary school and university students has been developed \cite{wolfbrenningwirkungschuelerin, wolfbrenningwirkungstudentin}. 
Strategic steps, followed by practical recommendations, are outlined within four main phases: process design, decision on questions and answers for measurement of success, questionnaire development, and data evaluation. 

There is evidence in the literature that the diversity of the student group should be considered when planning actions. Some actions have been shown to be  successful only with top performing students \cite{breda2020gender, fisk2021increasing} or with motivated students \cite{sobral2021flipped}. Others show negative impact on male students \cite{latulipe2018longitudinal, onyema2021effect}. 
In terms of encouraging women to pursue creative IT courses, e.g., game development, there is evidence that  women tend to choose subjects they have previously had experience with \cite{sigurdhardottir2019empowering}. It is therefore important to get them exposed early on.
Finally, research shows that once the initiatives stop, the numbers of women can drop again \cite{lagesen2021inclusion}. Hence, it is crucial to maintain continuous efforts to improve female participation in computing disciplines.

\section{CONCLUSIONS}
\label{conclusions}

While there appears to be no universal cure to address the ``leaky pipeline'', this research provides a comprehensive review of actions taken in a range of environments to enhance recruitment and retention of women in computing disciplines. Actions have been collated from a broad range of sources and categorised into four groups which are discussed in detail. 
Categorisation of gender balance actions is critical, providing  technical higher education organisations and indeed any stakeholders in STEM with a common language and understanding around action types and scale which will facilitate discussion and comparison of gender balance actions. 
Our review also examines practices around measuring the impact of gender balance actions. We identify the challenges of evaluating the impact of these actions. This work leads to our next project, the development of a toolkit of actions, with a mechanism for evaluation, to promote the recruitment and retention of women in undergraduate computing disciplines. 

\section*{ACKNOWLEDGEMENTS}
This project is funded by the Higher Education Authority of Ireland and Huawei Ireland. 



\bibliographystyle{unsrt} 
\bibliography{bibfile}



\end{document}